# Regenerative Soot-I: Carbon cluster formation in regenerative sooting plasmas

Shoaib Ahmad

*National Centre for Physics, Quaid-i-Azam University Campus, Shahdara Valley, Islamabad, 44000, Pakistan*

Email: sahmad.ncp@gmail.com

## Abstract

Laboratory formation of large carbon clusters $C_m$ (m ≤ $10^4$) in carbonaceous plasmas has been studied by using an especially designed ion source. Carbon is introduced into the glow discharge plasma by sputtering of the graphite electrode. Soot dominated plasma is created whose constituents are carbon clusters. It produces and recycles cluster containing plasma. Regenerative sooting plasma creates the environment in which the entire spectrum of clusters that contain the linear chains, rings and fullerenes. Velocity spectra of the extracted clusters have been measured with an ExB filter. These spectra indicate and identify the mechanisms operating in the soot.

---------------------------------------------------------------------------------------------------------

During Experiments with regenerative sooting plasmas [1] we have observed that carbon cluster $C_m$ synthesis is the most prominent carbon accretion mechanism in the glow discharge of a graphite hollow cathode. We have attempted to create conditions similar to those found in typical carbon stars [2,3] leading to *C* synthesis into structures ascribed to be spherical graphite grains. In most cool stars pressures of the order of $10^{-1}$–$10^{-5}$ mbar and temperature ~ 2000K at which partial pressures of carbon vapor is higher than the saturated vapor pressure of graphite [4]. Therefore, it has been suggested [5] that the precipitation of carbon atoms into graphite becomes possible. The starlight extinction hump around λ2175Å has been ascribed [5–8] to the presence of spherical graphite dust particles. Attempts have been made at fitting distributions of graphitic grain radii to explain the famous UV extinction hump [8]. Formation of soot that leads to the synthesis of $C_{60}$ has been used to explain this extinction by Kroto [9]. Our experiments suggest that the dominant mechanism for carbon clustering to occur in carbonaceous plasma is the formation of large $m$ ($\geq 32$) structures. We extract these carbon clusters $C_m$ in the formative as well as the well sooted stages of the so formed sooting plasmas. The $C_m$ spectra from the sooting plasma contain > 95% $C$ atoms in the range $m > 32$. The large cluster dominated plasma starts to produce fragments i.e. $m < 32$ if allowed to expand into an expansion chamber. Passing the same plasma through canals with axial magnetic field $B_0$ does not have a significant influence on the existing cluster distribution in well sooted plasma. On the basis of our experimental results we propose that the cusp field hollow cathode plasma can act as a source of regenerative sooting environment for the study of the evolution of carbon clusters. We may be able to study the energetics and the subsequent dynamics of clusters in environments similar to those existing in our source.

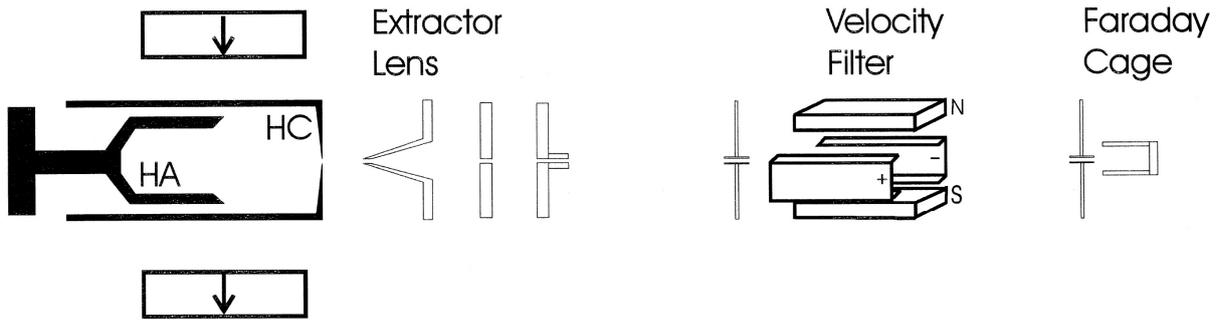

**Fig. 1.** The complete experimental setup is shown with the cluster ion source, extraction lens, collimators, the velocity filter and a Faraday cage. The source is composed of a Hollow Cathode (HC), Hollow Anode (HA) and a set of hexapole bar magnets.

A range of carbon cluster $C_m$ formation techniques has been developed [10 –15]. Laser ablation of graphite followed by supersonic expansion [10,11] yield the entire range of $C_m$ from $m = 11$ to $\geq 10^3$. Specific skimmer designs have preferential yields for fullerenes $C_{60}$ and $C_{70}$. High pressure arc discharge [12] is also a specialist technique for $C_{60}$ and $C_{70}$. Electron microscopy of high energy electron bombard soot [13] has established the existence of the shelled or hyper-fullerenes. Energetic ions have also been tried to produce closed cage structures in graphite and carbon containing condensed media [14–16]. Ion irradiations with keV [14] and MeV [15,16] energies have demonstrated the production of $C_m$ that include the linear chains, rings and fullerenes.

We report results of an investigation into the dynamics of soot formation from sputtered carbon atoms and ions in magnetized plasma. Fig. 1 shows the schematic diagram of the complete experimental setup. The source is shown in cathodic extraction mode. The extraction of plasma species from the source is done by a set of electrodes that extract as well as focus the beam. A set of collimating apertures restricts the beam divergence to within ±1 Å for analysis into a permanent magnet based velocity filter that has analyzing field ≈ 0.35 T on the axis. By varying the compensating electric field $\varepsilon_0$ the velocity analysis based on $v_0 = \varepsilon_0/B_0$ is performed with $m_0 \propto (\varepsilon_0)^{-1/2}$. A velocity spectrum always contains all velocities from $v_{min} \approx 0 \ (\equiv m_{max})$ to $v_{max}(\equiv m_{min})$. Dispersion $D$ of masses $m \pm \delta m$ is given by [17] as $D = al(\delta m/m)$. $\varepsilon_0/V_{ext}$ where $a$ and $l$ are the lengths of the velocity filter and the flight path, respectively. The extraction voltage $V_{ext}$ fixes the particle velocity for a given $B_0$. The velocity filter can be manipulated with the 3 independent variables $a$, $l$ and $V_{ext}$ to enhance the resolution of the low $v_0$ i.e., high $m$ clusters. Fig. 10 shows a specially designed hollow cathode ion source [1] comprising a graphite hollow cathode (HC), hollow anode (HA) and a hexapole bar magnet arrangement providing a maneuverable 3D cusp field geometry. The cusp field $B_z(r,\theta)$ contours are formed by the superposition of 2D fields $B_z(r)$ and $B_r(\theta)$. The source operates in the glow discharge mode at neon pressures $P_g \approx 10^{-1}$-$10^{-3}$ mbar and discharge voltage $V_d$ = 0.5 - 1.0 kV. Cathode wall sputtering by Ne$^+$ ions has been used to introduce $C$ into the plasma. The glow discharge has gradually increasing carbon content and it transforms into a sooting plasma that provides multistep clustering ($\sum_{m \geq 2} C_m \to C_m$) environment. The condensing carbon vapor from this sooting plasma onto the cathode walls deposits increasing layers of soot in two distinct processes: (1) Ne$^+$, C$^+$ and $C_m^+$ ions with energies ≈ 0.6 ± 0.2 keV collide with walls leading to secondary electron emission, sputtering and cluster disintegration. (2) Neutral and excited $C_m^{0,*}$ diffuse out of the plasma at their appropriate diffusion rates and may be adsorbed or reflected back. Sequences of deposition and subsequent removal by sputtering and thermal desorption of these layers leads to the formation of dynamic soot up to 40 ± 10 Å. This limit to soot layers is set by the range of the lightest ion species with energy ≈ 0.6 ± 0.2 keV. The sooting plasma-sooted wall interactions are the main mechanisms of soot regenerative sequences. The clustering properties of the source depend upon $B_z(r,\theta)$ the electrode design and the discharge parameters $V_d$, $i_d$, $P_g$.

Fig. 2a is the velocity spectrum of clusters $C_m^+$ from Ne sooted source and obtained after 100 min of operation at $V_d$ = 0.85 kV, $i_d$ = 100 mA and $V_{ext}$ = 1 kV. The spectrum is evenly spread between $C_m$ with 23 ≥ m ≥ 2 and those with m > 32. $C_1$ peak is the dominant one and clustering mechanisms seem to be operating.

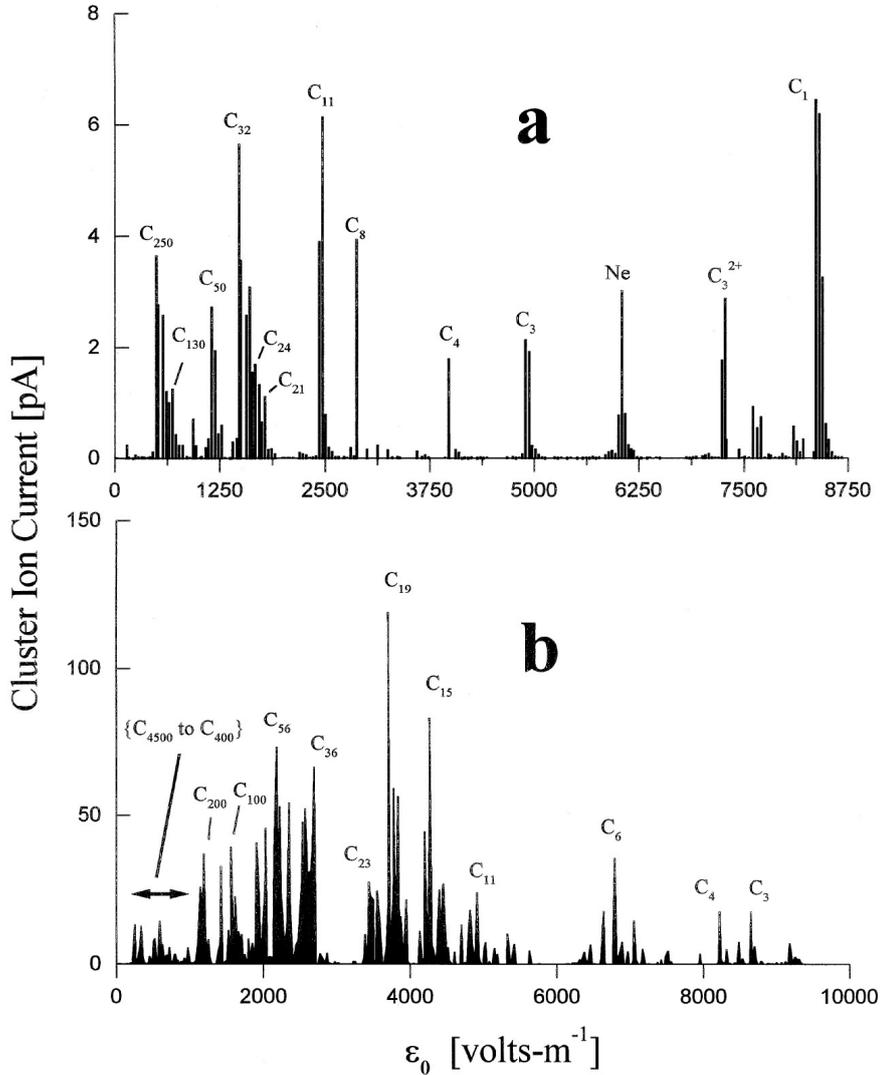

**Fig. 2.** Velocity spectra with $V_d$ = 0.85 kV, $I_d$ = 100mA and $V_{ext}$ = 1 kV. The spectrum is composed of the entire range of clusters. The multiply charged species can also be seen like $C_3^{2+}$. (b) $V_d$ = 0.5 kV, $I_d$ = 50mA and $V_{ext}$ = 2 kV. Well sooted source operated with Ne at ≈ 60W for 20 hours and then ignited with Xe. Note the cluster ion intensities difference in the mildly sooted (a) and the heavily sooted (b) source and the heavy mass range from $C_{4500}$ to $C_{400}$.

Fig. 2b on the other hand, is from the source sooted on Ne for 3 hr and then ignited with Xe. Extraction voltage $V_{ext}$ = 2 kV to separate higher masses (i.e., low values of $v_0$). The spectrum has $C_m$ in the increasing velocity order; firstly, the lowest velocity i.e., the heaviest clusters with $4300 \geq m > 300$ appear then the regime of clusters with $300 \geq m \geq 36$ and finally we have the smallest clusters with $23 \geq m \geq 3$. The cluster formation as well as fragmentation is continually taking place with the former being the dominant mechanism. The source operation at high $V_d$ produces heavier cluster dominated velocity spectra whereas, experiments with prolonged operation at low power input ($V_d \cdot i_d \leq 50W$) tend to stabilize plasma constituents and the resulting soot on the cathode.

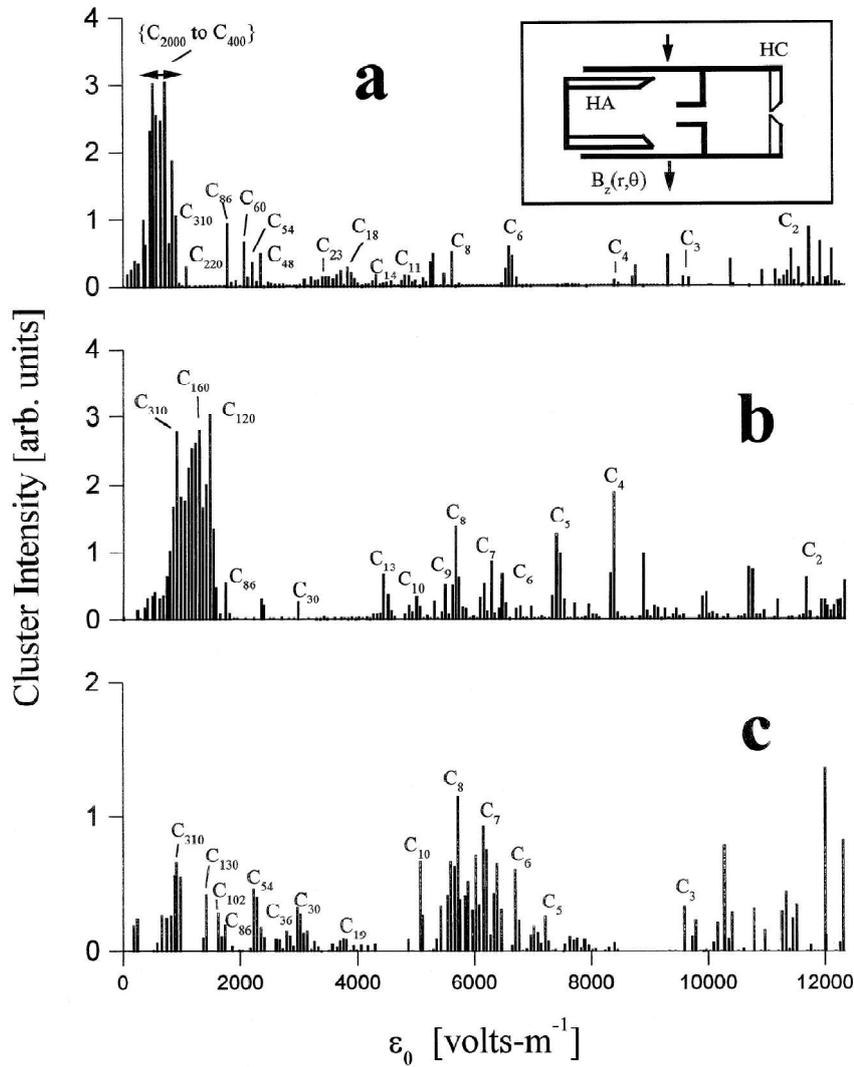

**Fig. 3.** Well sooted source plasma allowed expanding into an expansion chamber shown in the inset. Three consecutively collected velocity spectra show the re-bonding sequences leading to the growth of smaller clusters m < 10 at the expanse of the initially present heavy clusters.

If the sooted plasma is allowed to expand out of a 5mm aperture into an expansion chamber of 15mm diameter and 30mm length, the inter-cluster interactions and those of the clusters with charged and excited species like $Ne^*, C^*$ lead to the re-bonding/re-constitution of clusters. Fig. 3 is from the source that has been well sooted by operating for ~ 300 min with moderate power input i.e., $V_d \cdot i_d \approx 50W$. We present three spectra taken at half hourly interval with $V_d = 0.6$ kV, $i_d = 75$ mA, $P_g \sim 10^{-3}$ mbar and $V_{ext} = 2$ kV. In Fig. (3a) the first and most pronounced peak structure is within the range $C_{2130\pm500} \geq C_m \geq C_{380\pm20}$. Notice the error in determining the very heavy clusters. Then follows the C-cluster range from $C_{86}$ to $C_{48}$ and $C_m$ from $m \sim 23$ to 3. Within the peaks of $C_m$ ($23 \geq m \geq 3$) there are doubly or even higher charged clusters $C_m^{n+}$ ($n > 2$). These may arise from ionization in the extraction/focusing region by the back streaming electrons. Otherwise, velocity filter is not charge selective. From Fig. 3b and 3c the gradual diminution of large $m$ clusters and the growth of the smaller ones is observable. Here in Fig. 3 we have witnessed the recycling of the cathode soot and the velocity spectra indicated the regenerative processes of the sooting plasma while passing through an expansion chamber. Heavier cluster fragmentation favors the growth of the regime of small $m$ clusters.

In Fig. 4 one begins with a sooting plasma containing clusters that are predominantly of large $m$ (> 32). The plasma is allowed to expand into a narrow canal of 5mm diameter and 20mm length with an axial magnetic field $B_0 \sim 1.5$ T. Fig. 4a shows the preferential buildup of $C_{23}$, $C_{32}$, $C_{38} \rightarrow C_{54}$, $C_{78}$, $C_{102}$, $C_{120}$ and larger ones in the range $C_{3500}$ to $C_{300}$. The next spectrum of Fig. 4b has $C_{24}$ as the most significant peak. In addition, $C_{44}$, $C_{76}$, $C_{96}$, $C_{166}$, $C_{230}$ have well defined structures. Our mass resolution for heavier masses is poor at $V_{ext} = 1$ kV, but even with the inclusion of $\pm \delta m$ in the peak determination, the spectrum indicates that the heavy mass regime of clusters is stable. This is clearly evident from Fig. 4c that has $C_{60}$, $C_{102}$, $C_{260}$ as the most prominent peaks. $C_{24}$, $C_{28}$, $C_{38}$, $C_{76}$, $C_{166}$, $C_{260}$ are also present along with the broad continuous mass range $C_{1800}$ to $C_{300}$. Another noticeable feature is the low intensity of the smaller clusters in Fig. 4c.

Their relative peak heights are much less pronounced compared with their counter parts in Fig. 3. In this figure we have seen that the passage of sooting plasma through a restrictive canal super-imposed by an axial magnetic field has a high percentage of neutral, excited and ionized Ne and C species provide a stabilizing environment for the large $m$ cluster regime. This axial field is most effective with $C_1$, $C_2$, $C_3$ etc., therefore, one has a reservoir of small clusters gyrating along the axis of the canal that can not only compensate the loss of fragmenting clusters in subsequent collisions but also help the growth of heavier ones with $C_1$ and small clusters.

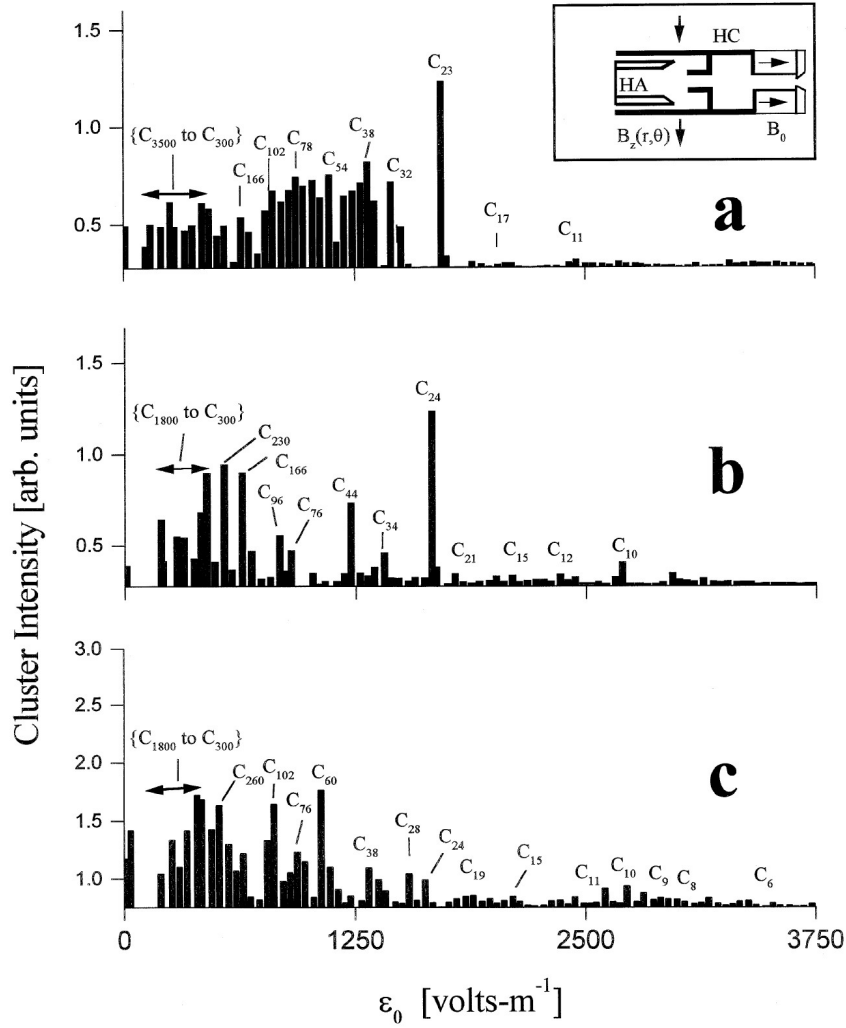

**Fig. 4.** Sooting plasma now expanding through a 5mm diameter, 20mm length canal with an axial magnetic field $B_0$. The inset shows the field configuration. The three successive spectra show the stable $C_m$ structures with some selected peaks around $C_{44}$, $C_{76}$, $C_{96}$, $C_{166}$, $C_{230}$ and the continuous higher mass range.

In this communication we have presented the results from a regenerative sooting source operating in the glow discharge mode. Sooting ensures large cluster formation in an environment of the sputtered $C$ vapour from the cathode walls. When the sooting plasma is allowed to expand into an expansion chamber large cluster fragmentation into smaller ones takes place. On the other hand, passage of the same expanding soot plasma through an axial magnetic field region does not produce cluster fragmentation. We propose, on the basis of our experimental results, that the formation of large carbon clusters in comparable carbonaceous environments of a lab scale hollow cathode sooting plasma can be compared with that of a cool carbon star. The most appropriate phase transformation of carbon vapor to 3D nano-structures may be passing through the graphite grain formation stages [5] or the evolution of fullerenes and the onion structures [9]. More experiments and varied techniques are needed to establish the structures of the observed large clusters formed in sooting plasmas.